\documentclass{article}

\usepackage[final]{neurips_2024}

\usepackage[utf8]{inputenc} 
\usepackage[T1]{fontenc}    
\usepackage{hyperref}       
\usepackage{url}            
\usepackage{booktabs}       
\usepackage{amsfonts}       
\usepackage{nicefrac}       
\usepackage{microtype}      
\usepackage{xcolor}         
\usepackage{graphicx}
\usepackage{times}
\usepackage{latexsym}
\usepackage{amssymb}
\usepackage{hyperref}
\usepackage{url}
\usepackage{booktabs}
\usepackage{color}
\usepackage{xcolor}
\usepackage{markdown}
\usepackage[most]{tcolorbox}
\usepackage{multirow} 
\usepackage{tabularx} 
\usepackage{booktabs} 
\usepackage{array}    
\usepackage{longtable}  
\usepackage{pgfplots}
\usepackage{float}
\usepackage{IEEEtrantools}
\usepackage{amsmath}
\usepackage{titlesec}
\usetikzlibrary{patterns}

\title{Audio-FLAN: An Instruction-Following Dataset for Unified Audio Understanding and Generation of Speech, Music, and Sound}
\author{
    \IEEEauthorblockN{Liumeng Xue$^{a,b*}$, Ziya Zhou$^{a,b*}$, Jiahao Pan$^{a,b}$}
    \IEEEauthorblockN{\textbf{Zixuan Li$^{c}$, Shuai Fan$^{d}$, Yinghao Ma$^{e,b}$, Sitong Cheng$^{a}$}}
    \IEEEauthorblockN{\textbf{Dongchao Yang$^{f}$, Haohan Guo$^{f}$, Yujia Xiao$^{f}$, Xinsheng Wang$^{a}$}}
    \IEEEauthorblockN{\textbf{Zixuan Shen$^{a}$, Chuanbo Zhu$^{a}$, Xinshen Zhang$^{a}$, Tianchi Liu{$^g$}}}
    \IEEEauthorblockN{\textbf{Ruibin Yuan$^{a,b}$, Zeyue Tian$^{a,b}$, Haohe Liu$^{b,h}$}}
    \IEEEauthorblockN{\textbf{Xingjian Du$^{i}$, Emmanouil Benetos$^{b,e}$, Ge Zhang$^{b}$}}
    \IEEEauthorblockN{\textbf{Xu Tan$^{j}$, Yike Guo$^{a}$, Wei Xue$^{a}$}}
    \vspace{0.1in}
    \IEEEauthorblockA{$^a$ The Hong Kong University of Science and Technology, $^b$ M-A-P}
    \IEEEauthorblockA{$^c$ Inner Mongolia University, $^d$ Beihang University}
    \IEEEauthorblockA{$^e$ Queen Mary University of London, $^f$ The Chinese University of Hong Kong}
    \IEEEauthorblockA{$^g$ National University of Singapore, $^h$ University of Surrey, }
     \IEEEauthorblockA{$^i$ University of Rochester, $^j$ Independent Researcher}
}

\AtBeginDocument{
  \newgeometry{
    textheight=9in,    
    textwidth=6.5in,   
    top=1.5in,         
    bottom=1.5in,      
    left=1in,          
    right=1in,         
    headheight=12pt,   
    headsep=25pt,      
    footskip=30pt      
  }
}
\begin{document}

\maketitle
\begin{abstract}
Recent advancements in audio tokenization have significantly enhanced the integration of audio capabilities into large language models (LLMs). However, audio understanding and generation are often treated as distinct tasks, hindering the development of truly \textbf{unified} audio-language models. While instruction tuning has demonstrated remarkable success in improving generalization and zero-shot learning across text and vision, its application to audio remains largely unexplored. A major obstacle is the lack of comprehensive datasets that unify audio understanding and generation. To address this, we introduce \textbf{Audio-FLAN}, a large-scale instruction-tuning dataset covering 80 diverse tasks across speech, music, and sound domains, with over 100 million instances. Audio-FLAN lays the foundation for unified audio-language models that can seamlessly handle both \textbf{understanding} (e.g., transcription, comprehension) and \textbf{generation} (e.g., speech, music, sound) tasks across a wide range of audio domains in a zero-shot manner. The Audio-FLAN dataset is available on HuggingFace~\footnote{\url{https://huggingface.co/HKUSTAudio}} and GitHub~\footnote{\url{https://github.com/lmxue/Audio-FLAN}}.

\end{abstract}
\section{Introduction} 

Recent advances in large language models and multimodal models have highlighted the effectiveness of \emph{instruction tuning} for broad generalization~\citep{ouyang2022traininglanguagemodelsfollow,touvron2023llama,achiam2023gpt}. Instruction-tuned models can generalize to unseen tasks far better than task-specific counterparts. In the text domain, models like FLAN (Finetuned Language Net)~\citep{wei2021flan} demonstrate remarkable zero-shot and few-shot capabilities when fine-tuned on diverse instructions. For example, FLAN (137B parameters) was fine-tuned on 60 NLP tasks and outperformed even larger models, like the 175B GPT-3~\citep{brown2020gpt3}, on many unseen tasks. Similarly, LIMA~\citep{zhou2024lima}, which used only 1,000 curated examples, achieved results preferred over much larger models, showing that minimal high-quality instruction data can significantly improve a model's ability to follow complex queries. In the vision domain, unified models like Chameleon~\citep{team2024chameleon} and Janus-Pro 7B~\citep{Janus} have demonstrated strong performance by handling both understanding and generation tasks in a single system, outperforming specialized models in image captioning, visual question answering, and image generation. In contrast, the audio domain~\footnote{In this paper, `audio' refers to two distinct meanings: (a) in a narrower sense, `audio' refers to `sound', which is related to but different from speech and music, often used in the context of `speech, music, and audio'; (b) in a broader sense, `audio' encompasses speech, music, and sound, used in the context of `text, vision, and audio'.} still lags behind, with audio understanding and generation often treated as separate tasks.

This gap between modalities highlights a critical limitation: \textbf{audio-language models still lack the unified modeling and generalization capabilities} that are now common in NLP and computer vision. Despite the wide variety of audio tasks (such as speech transcription, speaker identification, emotion recognition, sound event recognition, music understanding, and text-to-speech generation), there is no "audio GPT" or "audio foundation model" that can seamlessly switch between understanding and generating audio across speech, music, and audio domains. For example, models like Musilingo~\citep{deng2023musilingo} focus on music understanding, while LTU (Listen, Think, Understand)~\citep{gong2023listen} and Audio-Flamingo~\citep{kong2024audioflamingonovelaudio} focus on the audio domain. The SALMONN~\citep{tang2023salmonn} and Qwen-Audio series~\citep{chu2023qwen} are designed for understanding speech, sound, and music, but lack generation capabilities. On the other hand, UniAudio~\citep{yang2023uniaudio} supports audio generation, but it is limited to 11 tasks spanning speech, sound, music, and singing, each with specific task identifiers.

Currently, no audio model exhibits the broad zero-shot generalization seen in text and vision models. Recent benchmarks highlight these limitations. Dynamic-SUPERB~\citep{huang2024dynamicsuperbdynamiccollaborativecomprehensive}, a comprehensive benchmark with 33 speech tasks for speech models, shows that unlike text models, speech models remain confined to narrow tasks. It finds that systems perform well on seen tasks but struggle with unseen tasks, revealing poor zero-shot generalization. Dynamic-SUPERB Phase-2~\citep{dynamicsuperb2}, which has expanded to include 180 understanding tasks, reports that while recent models perform well on specific tasks, they struggle with generalization, underscoring the need for more research on developing universal models. Similarly, the MMAU benchmark~\citep{mmau}, which covers speech, environmental sounds, and music, shows that even top models like Gemini-Pro v1.5~\citep{team2024gemini} and Qwen2-Audio~\citep{chu2024qwen2} only achieve about 52\% accuracy. This stark contrast with text models underscores the underexplored potential of audio-language models for general auditory intelligence. Additionally, the lack of comprehensive evaluation frameworks further hinders progress. AIR-Bench~\citep{AIR_Bench}, the first generative audio-language comprehension benchmark, reveals significant limitations in current models' ability to follow instructions across tasks. In summary, audio-language research is still in an early stage, similar to the pre-GPT-3/FLAN era of NLP: while there are task-specific models, there is no unified model with broad, zero-shot capabilities.

A key challenge in the audio domain is \textbf{the lack of large-scale, diverse instruction-tuning datasets tailored to audio-language tasks}. While NLP has benefited from extensive multi-task instruction datasets like Super-NaturalInstructions~\citep{Super_naturalinstructions} with 1,616 tasks and vision-language models use resources like LLaVA~\citep{LLaVA} and InstructBLIP~\citep{instructblip}, the audio field lacks comparable datasets in scale or diversity. 
Some efforts, like GAMA~\citep{GAMA} synthesize an instruction dataset, called CompA-R, for audio reasoning, but they focus mainly on narrow tasks like question-answering and captioning. Other works have used GPT-4 or LLMs to generate instruction data from existing speech corpora,  e.g., LTU~\citep{gong2023listen} and LTU-AS~\citep{gong2023joint}, but these are fragmented, limited in scope, and often biased by the prompts used. 
No existing dataset spans the breadth of audio content, including speech, music, and sound, with instructions. In short, the audio domain lacks a “FLAN” equivalent—a consolidated, high-quality instruction dataset to unify myriad audio tasks. This absence of data is a key reason we do not yet have audio models with the generalization of GPT-4 or Chameleon. Even as benchmarks like the Dynamic-SUPERB series and AIR-Bench call for instruction-following audio models, researchers struggle to train such models without a large, diverse training corpus tailored to audio-language understanding and generation.

In this work, we introduce \textbf{Audio-FLAN}, a preliminary attempt to bridge this data gap and enable truly unified audio-language modeling. Audio-FLAN (Preliminary Release) is \textbf{a large-scale, diverse instruction-tuning dataset for both understanding and generation tasks across speech, music, and audio}, constructed by collecting and standardizing nearly all publicly available academic audio datasets into a common instruction-based format. By normalizing the format of these heterogeneous datasets, we provide each audio sample with one or more accompanying instructions (or question/prompt) and the expected output (transcription, description, answer for understanding tasks, or an audio clip for generative tasks). Crucially, Audio-FLAN is designed to support both pre-training and supervised fine-tuning (SFT) of models for unified audio-language tasks. We envision that models trained on Audio-FLAN dataset will be capable of both audio understanding (e.g., transcribing and comprehending audio, answering questions about it) and audio generation (e.g., following instructions to produce speech, music and sounds) within one unified framework. In other words, Audio-FLAN lays the groundwork for an audio equivalent of multimodal foundation models—an audio-language model that can listen, understand, speak, sing and compose in a general way.

To our knowledge, \textbf{Audio-FLAN} is the first comprehensive compilation that combines diverse audio datasets into a single, instruction-driven corpus of considerable scale. It includes approximately \textbf{80 tasks} and over \textbf{100 million} instances, significantly surpassing prior efforts in both quantity and diversity. We aim for Audio-FLAN to achieve for audio what FLAN and other instruction-tuned models have accomplished for text—enabling models to generalize across a wide range of audio tasks in a zero-shot manner and follow open-ended instructions related to audio content. The preliminary release of Audio-FLAN is only the beginning: we invite the research community to build on this resource, contribute new tasks (similar to Dynamic-SUPERB Phase-2), and explore unified models for speech, music, and audio. By unifying both audio understanding and generation, Audio-FLAN paves the way toward foundational models that can hear and generate audio as flexibly and broadly as language models process text.

\section{Audio-FLAN Dataset Construction}





\begin{figure*}[thbp]
\centering
\includegraphics[width=\textwidth, height=\textheight, keepaspectratio]{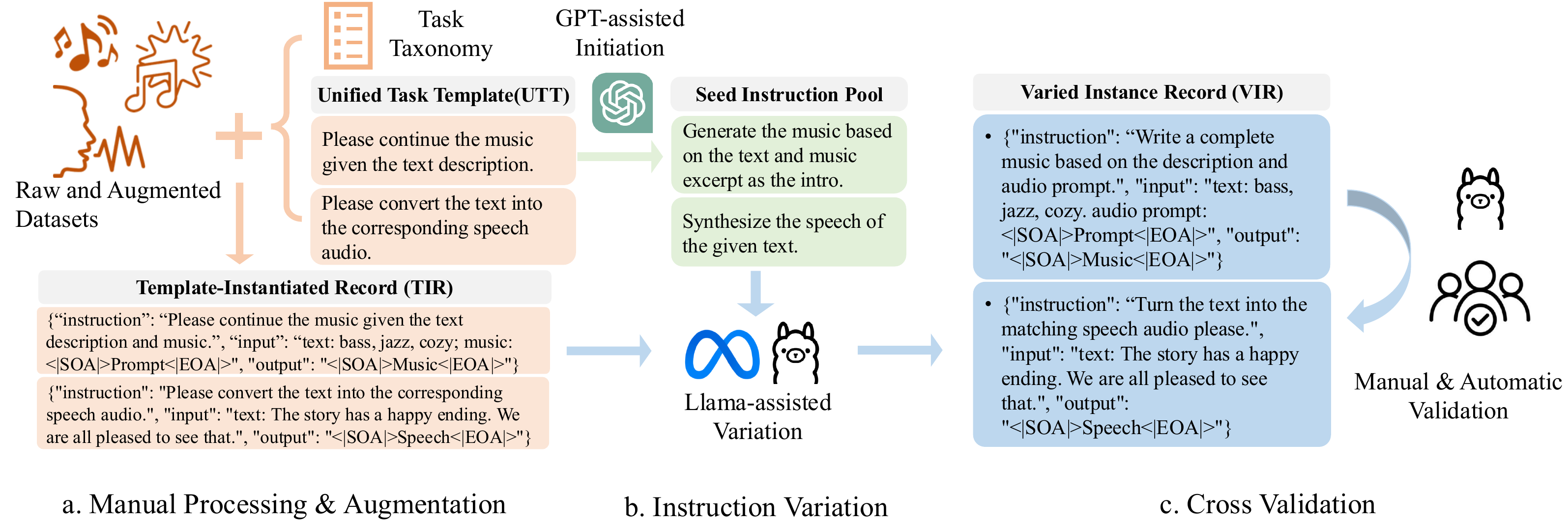}
\caption{Overview pipeline of Audio-FLAN dataset construction.}
\label{fig:dataset_pipeline}
\end{figure*}

Figure~\ref{fig:dataset_pipeline} illustrates the pipeline for constructing the Audio-FLAN dataset. We first collect the publicly released datasets and use their original labels, or manually processed labels, to determine the tasks that can be performed based on the task definitions. Next, instructions are generated and structured using task templates, which guide the format and content of instruction, input, and output. To increase the diversity of the instruction set, we apply a self-instruct-like method~\citep{wang2023self}, where the instructions are varied through tools like LLaMA and GPT, which allow for the creation of multiple variations for each task and instance. These varied instructions are then validated to ensure they meet the required standards before being integrated into the dataset.

\subsection{Task Category}
We classify tasks into \textbf{Major Tasks} and \textbf{Minor Tasks} following a hierarchical structure based on the scope and specificity of the tasks within the broader domains of \textbf{speech}, \textbf{music}, and \textbf{audio}, as shown in Table~\ref{tab:major_minor_task}. 
\begin{itemize}
    \item \textbf{Major Tasks} represent broad categories that encompass a variety of related activities within each domain. For example, in the speech domain, major tasks include \textit{Speech Recognition}, \textit{Speech Generation}, and \textit{Speech Enhancement}, which cover the general areas of recognizing spoken words, generating speech, and improving speech quality, respectively.
    \item \textbf{Minor Tasks} are specific subcategories under each major task, providing more focused and detailed areas of work. For example, under \textit{Speech Recognition}, the minor tasks include \textit{Automatic Speech Recognition}, \textit{Dialect Automatic Speech Recognition}, and \textit{Phonetic Recognition}, each representing a specialized area within the overarching task of recognizing speech. Similarly, under \textit{Speech Generation}, tasks like \textit{Text to Speech}, \textit{Voice Conversion}, and \textit{Speech to Speech Translation} address more specific aspects of generating speech.
\end{itemize}



\begin{longtable}{p{1.5cm}p{6cm}p{7cm}}   
\caption{Task category in Audio-FLAN dataset.}  
\label{tab:major_minor_task}  
\\  
\hline
\textbf{Domain} &  \textbf{Major Task} & \textbf{Minor Task}  \\
\hline
\endfirsthead  
\multicolumn{3}{|c|}{\textit{Continued from the previous page}} \\  
\hline
\textbf{Domain} &  \textbf{Major Task} & \textbf{Minor Task} \\
\hline
\endhead  
\hline
\endfoot  
\hline
\endlastfoot  
    Speech & \multirow{3}{*}{Speech Recognition} & Automatic Speech Recognition          \\  %
                             \multirow{33}{*}{} &                      & Dialect Automatic Speech Recognition                 \\  %
                             &                      & Phonetic Recognition                                 \\  %
                             \cline{2-3} %
                            & \multirow{2}{*}{Spoken Language Understanding} & Intent Classification          \\  %
                             &                      & Speech to Text Translation            \\  %
                             \cline{2-3} %
                            & \multirow{7}{*}{Paralinguistic Attribute Recognition} &  Gender Recognition      \\
                            &                                       &  Age Recognition                         \\
                            &                                       &  Emotion Recognition                     \\
                            &                                       &  Accent Recognition                      \\
                            &                                       &  Spoken Paragraph Recognition            \\
                            &                                       &  Language Identification                 \\
                            &                                       &  Dialect Identification                  \\
                            \cline{2-3} %
                            & \multirow{4}{*}{Speaker Recognition}   & Speaker Verification                   \\
                            &                        & Speaker Diarization           \\
                            &                        & Speaker Extraction            \\
                            &                        & Speaker Identification        \\
                            \cline{2-3} %
                            & \multirow{1}{*}{Speech Caption}    & Speech Caption        \\
                            \cline{2-3} %
                            & \multirow{3}{*}{Speech Detection}  & Deepfake Detection    \\
                                            &    & Vocoder Type Classification           \\
                                            &    & Device Recognitin           \\
                                             \cline{2-3}
                            & \multirow{5}{*}{Speech Enhancement}    & Denoising     \\
                            &                        & Dereverberation               \\
                            &                        & Declipping                    \\
                            &                        & Speech Bandwidth Extension        \\
                            &                        & Signal-to-noise Ratio Estimation  \\
                            \cline{2-3} %
                            & \multirow{9}{*}{Speech Generation}     & Text to Speech        \\  %
                            &                        & Zero-shot Text to Speech              \\  %
                            &                        & Emotional Text to Speech              \\  %
                            &                        & Zero-shot Emotional Text to Speech    \\  %
                            &                        & Descriptive Speech Synthesis          \\  %
                            &                        & Spontaneous Text to speech            \\  %
                            &                        & Voice Conversion                      \\  %
                            &                        & Emotion Conversion                    \\  %
                            &                        & Speech to Speech Translation          \\  %
                            \cline{2-3} %
     \cline{2-3}
    Total & 8 & 34  \\  
    \pagebreak
    Music & \multirow{10}{*}{Global MIR} & Key Detection \\
                           \multirow{27}{*}{} &             & Scale Recognition                    \\ 
                           &             & Music Tagging                        \\ 
                           &             & Genre Classification                 \\ 
                           &             & Emotion Classification               \\ 
                           &             & Pitch Classification                 \\ 
                           &             & Instrument Classification            \\ 
                           &             & Vocal Technique Classification       \\ 
                           &             & Instrumental Technique Classification \\ 
                           &             & Artist Identification \\ 
                    \cline{2-3} %
                         & \multirow{3}{*}{Sequential MIR} & Beat Tracking  \\  
                         &                 & Chord Estimation  \\
                         &                 & Progression Extraction \\
                     \cline{2-3} %
                         & \multirow{2}{*}{Single Music Reasoning} & Beat-level Instruments Recognition  \\  
                         &                         & Beat-level Pitch Estimation  \\  
                     \cline{2-3} %
                         & \multirow{5}{*}{Multiple Music Reasoning} & Tempo Comparison  \\
                         &                           & Instrument Comparison  \\
                         &                           & Key Comparison \\
                         &                           & Instrumental Technique Comparison \\
                         &                           & Emotion Comparison  \\
                     \cline{2-3} %
                         & Music Caption & Music Caption  \\
                     \cline{2-3} %
                         & \multirow{2}{*}{Music Separation}& Melody Extraction \\
                         &                 & Text-guided Source Separation \\
                     \cline{2-3} %
                         & \multirow{5}{*}{Music Generation} & Text-to-music Generation  \\
                         &                   & Text-guided Music Continuation   \\
                         &                   & Lyrics-to-song Generation           \\
                         &                   & Singing Voice Synthesis          \\
                         &                   & Singing Voice Conversion         \\
                         
     \cline{2-3}
     Total & 7 & 28  \\  
    \midrule
    Audio & \multirow{4}{*}{Audio Event Recognition} & Sound Event Sequence Recognition \\
                           \multirow{17}{*}{} &                                           & Sound Event Recognition  \\
                           &                                           & Sound Event Detection  \\
                           &                                           & Acoustic Scene Classification  \\
                           \cline{2-3} %
                           & Audio Caption & Audio Caption \\
                           \cline{2-3} %
                         & Audio Advanced Understanding & Sound Event Understanding  \\
                         \cline{2-3} %
                         &  \multirow{2}{*}{Audio Detection} & Deepfake Audio Detection \\
                         &                  & Voice Activity Detection  \\
                         \cline{2-3} %
                         & \multirow{2}{*}{Audio Classification} & Speech, Silence, Music and Noise Classification \\
                         &                                        & Speech Nonspeech Detection \\
                         \cline{2-3} %
                         & \multirow{2}{*}{Audio Enhancement} &  Audio Inpainting     \\
                         &                                   & Audio Super-resolution    \\
                         \cline{2-3} %
                         &  \multirow{3}{*}{Audio Separation} & Text-guided Audio Source Separation  \\
                         &                   & Label-querying Sound Extraction \\
                         &                   & Audio-querying Sound Extraction \\                        
                         \cline{2-3} %
                         & \multirow{3}{*}{Audio Generation} & Text-guided Audio Generation  \\
                         &                   & Time-grounded Text-to-audio Generation        \\
                         &                   & Audio Continuation  \\
                         \cline{2-3} %
     
     \cline{2-3}
    Total & 8 & 18  \\  
    \midrule
    \midrule
     \textbf{Total}& \textbf{23} & \textbf{80}  \\
    \bottomrule
\end{longtable}

This hierarchical approach provides a clear structure that allows for easy navigation of the tasks. By categorizing tasks into major and minor tasks, it is easier to understand the broad objectives as well as the specific challenges and techniques involved in each sub-area.
Besides, this classification system allows researchers and practitioners to target specific areas of interest. 
Furthermore, the system is flexible, accommodating new tasks as the fields evolve. New minor tasks can be added under existing major tasks, or new major tasks can be created as technology advances, ensuring that the classification system can adapt to future developments.

The \textbf{Audio-FLAN} dataset introduces time-sequential tasks that have been underexplored in previous research, particularly in the textual domain, as time sequences are a distinctive feature of the audio domain. These include tasks like \textit{Melody Extraction}, and \textit{Pitch Estimation} (with timestamps) in the music domain, as well as \textit{Sound Event Sequence Recognition} and \textit{Sound Event Detection} (with timestamps) in the audio domain. These tasks require processing entire audio sequences or segments, highlighting the importance of time-based analysis. In the speech domain, tasks like \textit{Spoken Paragraph Recognition} further emphasize the role of time sequences, as the model must compare recordings and analyze linguistic content aligned over time. 

Additionally, text-based LLMs are often praised for their reasoning capabilities in tackling complex tasks that involve interdependent results. In the music domain, we introduce reasoning tasks where models must first localize a time segment based on instructions and then perform estimations to generate precise answers. For example, \textit{Beat-level Pitch Estimation} and \textit{Beat-level Instrument Recognition} (under \textit{Single Music Reasoning}) require models to interpret musical elements at specific time points, while \textit{Tempo/Key/Instrument/Emotion Comparison} (under \textit{Multiple Music Reasoning}) involves comparing musical features over time. These tasks push the limits of model generalization across complex, time-based data, positioning \textbf{Audio-FLAN} as a unique resource for developing unified models capable of processing time-sensitive audio across speech, music, and audio.

In conclusion, the hierarchical classification system effectively organizes each domain into high-level tasks (Major Tasks) and more specific subtasks (Minor Tasks), providing a clear structure. With \textbf{23 major tasks} and \textbf{80 minor tasks}, the dataset covers a wide range of understanding and generation tasks across speech, music, and audio, underscoring the depth of research and application in these fields. Notably, the \textbf{Audio-FLAN} dataset is the first instruction-tuning dataset to incorporate tasks from \textbf{speech}, \textbf{music}, and \textbf{audio}, addressing both \textbf{generation} and \textbf{understanding} tasks. This contribution fosters the development of unified audio-language models with generalization capabilities similar to those in the NLP and computer vision domains.

\subsection{Dataset Processing}

Our goal is to develop a large and diverse instruction dataset by aggregating tasks from various domains and applications. Building such an extensive instruction dataset from scratch would be highly resource-intensive and time-consuming. To mitigate this challenge, we leverage existing audio datasets from the research community, transforming them into an instructional format. This approach capitalizes on the wealth of labeled data that is already available or manually processed, allowing us to repurpose datasets for broader applications. Specifically, we aggregate over 52 datasets that are either publicly accessible or can be obtained upon request. The datasets associated with each task are listed in Table~\ref{tab:task2dataset}.

In the speech, music and audio domains, many tasks depend heavily on \textbf{pre-labeled data}, such as genre labels, speech annotations, or musical characteristics. For instance, tasks like \textit{Automatic Speech Recognition (ASR)} and \textit{Text-to-Speech (TTS)} rely on paired text and speech data, while \textit{Emotion Recognition} and \textit{Gender Recognition} tasks in speech utilize emotion and gender labels, respectively. In the \textit{Music} domain, tasks like \textit{Genre Classification} and \textit{Emotion Classification} require labeled music data with genre or emotion tags, and \textit{Pitch Classification} and \textit{Instrument Classification} rely on instrument-specific annotations. 
However, there are several tasks for which suitable labeled datasets are not readily available or require additional processing. For example, tasks such as \textit{Audio Inpainting} or \textit{Music Generation} often lack directly available labels or training data that match the specific needs of these tasks. In these cases, \textbf{manual processing} is required to create the necessary data.


In the \textbf{speech} domain, for \textit{Speech Enhancement} tasks, data simulation techniques generate task-specific datasets from clean speech corpora. For \textit{Denoising}, noisy-clean pairs are created by adding noise to clean speech samples. \textit{Dereverberation} involves generating reverberant-clean pairs by convolving clean speech with real or simulated room impulse responses. In the \textit{Declipping} task, clean speech is randomly clipped for model input. For \textit{Speech Bandwidth Extension}, high-sample-rate speech is downsampled to teach the model how to recover high-quality speech from lower-quality input. In \textit{Speaker Recognition}, \textit{Speaker Extraction} creates datasets by mixing clean speech from multiple speakers and providing reference speech for the target speaker.


Similarly, the \textit{Music Generation} tasks in the \textbf{music} domain, such as for the \textit{Text-guided Music Continuation} or \textit{Lyrics-to-song Generation}, manually processed data might be needed to create the text-to-music pairs. This could involve taking existing music pieces and pairing them with relevant textual descriptions, or generating new musical content based on textual input using music generation models. In cases where music data is not paired with lyrics, data augmentation techniques might be used, where new synthetic music tracks are generated by modifying or extending the existing ones to suit the task.

In the \textbf{audio} domain, the \textit{Audio Generation} tasks such as \textit{Audio Inpainting}, the data processing involves selecting clean audio samples, cutting them to create gaps, and preparing the dataset for further use in reconstructing the missing segments. In \textit{Audio Super-resolution}, the process includes downsampling high-quality audio to a lower resolution and then using the downsampled version to recreate the original high-resolution audio. These processing steps facilitate the generation of suitable datasets for these tasks.


These cases highlight the flexibility and adaptability of existing datasets in the speech, music, and audio domains, where manual dataset processing and augmentation are crucial for handling tasks with limited labeled data or where the required labels do not exist. By applying these dataset processing techniques, we can ensure that tasks with scarce resources are still effectively addressed, broadening the applicability of existing datasets to more diverse machine learning applications. Furthermore, the \textbf{Audio-FLAN} dataset is continuously being expanded and processed to cover additional tasks. We also invite all interested researchers and practitioners to contribute to the ongoing development of the \textbf{Audio-FLAN} instruction tuning dataset, enhancing its scope and utility for the community.



\subsection{Task Instruction Template}
The instruction data we aim to generate consists of a collection of instructions \( \{I_{i}\} \), each describing a specific task \( i \) in natural language. For each task \( i \), there are \( n_{i} \geq 1 \) input-output pairs \( \{(X_{t,i}, Y_{t,i})\}_{t=1}^{n_{i}} \). Once the tasks to be covered by the dataset are determined, we process the data into three core components: \textbf{instruction}, \textbf{input}, and \textbf{output}, all formatted in JSONL (JSON Lines) format. The \textbf{instruction} serves as a concise description of the task, guiding the model on the expected input and the type of output to generate. For tasks that involve understanding, the \textbf{output} is \textit{text}, while for tasks focused on generation, the \textbf{output} is typically audio. The \textbf{input} can be \textit{audio}, \textit{text}, or a combination of both, depending on the task. Formally, given this structured data, a model \( M \) is expected to generate the appropriate output based on the task instruction and the corresponding input: \( M(I_{i}, X_{t,i}) = Y_{t,i}, \quad \text{for } i \in \{1, \ldots, n_{i}\} \).

In the \textbf{speech} domain, the task of \textit{Speech-to-Text Translation} involves both text and audio as input (e.g., an audio recording of speech and the corresponding transcription in target language), and the output is text, which is the translated text in a different language. In the \textbf{music} domain, the task of \textit{Text-guided Music Generation} uses a combination of text and audio as input (e.g., a description of the type of music and a short melody clip), and the output is audio, which is a generated music track that matches the input description and melody. In the \textbf{audio} domain, tasks like \textit{Audio Super-resolution} can take a combination of low-resolution audio and textual description of the expected quality improvements as input, and the output is high-resolution audio that enhances the quality of the input signal.

To generate the task instructions \( \{I_{i}\} \), we initially employ template-based instructions. These instructions are human-written, task-specific descriptions that explicitly define the task. For example, the \textbf{instruction} for the \textit{Speech-to-Text Translation} task could be "Please translate the speech into the text in Chinese.". For \textit{Text-guided Music Generation}, the \textbf{instruction} might be "Please continue the audio music prompt based on the given text description." The \textbf{instruction} for the \textit{Audio Super-resolution} task can be "Please increase the resolution of the given audio signal to 32K Hz". Here are the three task instruction templates:
\begin{tcolorbox}[colback=white, colframe=black, boxrule=0.2mm, arc=0mm, title=Speech-to-Text Translation]
\{
"instruction": "Please translate the speech into the text in English.", 
"input": "<|SOA|>Audio\_ID<|EOA|>", 
"output": "Nevertheless, there are many distinctive ways of drinking coffee around the world that are worth experiencing."
\}
\end{tcolorbox}


\begin{tcolorbox}[colback=white, colframe=black, boxrule=0.2mm, arc=0mm, title=Text-guided Music Continuation]
\{
"instruction": Please continue the audio music prompt based on the given text description",
"input": "This is a Carnatic music piece set in the atana raga. It follows the 5/8 meter and is composed in the khandaChapu taala. The lead instrument featured in this performance is vocal, accompanied by Mridangam. The kalai of this composition is 1. \textbackslash n audio prompt: <|SOA|>Audio\_ID<|EOA|>",
"output": "audio: <|SOA|Audio\_ID<|EOA|>"
\}
\end{tcolorbox}


\begin{tcolorbox}[colback=white, colframe=black, boxrule=0.2mm, arc=0mm, title=Sound Super-resolution]
\{
"instruction": "Please increase the resolution of the given audio signal to 32k Hz.", 
"input": "audio: <|SOA|>Audio\_ID<|EOA|>."
 "output": "<|SOA|>Audio\_ID<|EOA|>",
\}
\end{tcolorbox}


We include \texttt{<SOA>} to mark the start of audio, and \texttt{<EOA>} to signify the end of audio. When the input contains multiple values, they are separated by \texttt{\textbackslash n}. Note that the JSONL format files contain not only the \textit{instruction}, \textit{input}, and \textit{output}, but also other relevant fields such as \texttt{uuid}, \texttt{split}, \texttt{task\_type}, and \texttt{domain}. The complete JSON file content can be found in Appendix~\ref{appendix:instruction_template}. These task-specific templates serve as foundational structures, which can later be refined and expanded upon to better suit a wide range of tasks across different domains. This method ensures that the instructions are both clear and aligned with the model's input-output expectations.

\subsection{Instruction Variation}\label{sec:self_instruct}
While fixed, template-based instructions provide consistency in task execution, they inherently constrain flexibility and creativity. This rigidity can hinder the model’s ability to adapt to diverse and nuanced task descriptions. To mitigate these limitations and enhance the diversity and creativity of the instructions, we introduce an approach that expands template-based instructions into a broader set of variations using advanced language models, like LLaMA~\citep{touvron2023llama}. By leveraging the generative power of these models, we can produce multiple distinct variations for each task instruction template, thereby augmenting the model's capacity to handle a wide array of task descriptions.

The process of instruction variation follows a three-step pipeline, inspired by the self-instruct approach~\citep{wang2023self}, designed to systematically enhance instruction diversity. These steps include: (1) initializing the variation seed pool, (2) generating new diverse instructions, and (3) validating the generated instructions.

In the first step, we begin by generating five new instruction examples for each task using GPT-4o, which serves as the initial "seed" pool. These initial variations form the basis for subsequent instruction generation. In the second step, we utilize the Llama-3.1-70B-Instruct model to generate instruction variation, drawing from the seed pool. Llama-3.1-70B-Instruct allows for the generation of diverse and contextually varied instructions, along with modifying or adding prefixes within the \textit{input} and \textit{output} fields based on the specific characteristics of the task. This process allows for further customization of task instructions that are both rich in variation and contextually appropriate.

The final step involves rigorous validation of the generated instructions to ensure their integrity and quality. Specifically, we verify that the audio ID remains consistent with the original task instance and confirm that the JSONL format adheres to the required structure. Any variations that exhibit formatting errors, such as incorrect JSONL syntax or mismatched audio IDs, are identified and excluded from the pool. Any instructions deemed invalid are flagged for regeneration, and if no suitable variation can be generated by the model, manual intervention is employed to address the issue. This ensures that both the quantity and quality of the variations are maintained. Valid instructions are then reintegrated into the task pool for use in generating further variations. 

This iterative process promotes a dynamic and evolving pool of task instructions, effectively maximizing their diversity. As a result, the model becomes more adept at handling a wide range of task descriptions, ultimately improving its overall performance and generalization ability across diverse use cases. 
The prompt used to produce various instructions by GPT-4 and LLaMA is provided in Appendix~\ref{appendix:instruction_variation_prompt}. Specific examples of the instruction template and generated instruction variations are shown in Appendix~\ref{appendix:instruction_variation_example}.



\section{Audio-FLAN Dataset}

\begin{figure*}[h] 
\centering
\scalebox{0.9}{
\includegraphics[width=\linewidth]{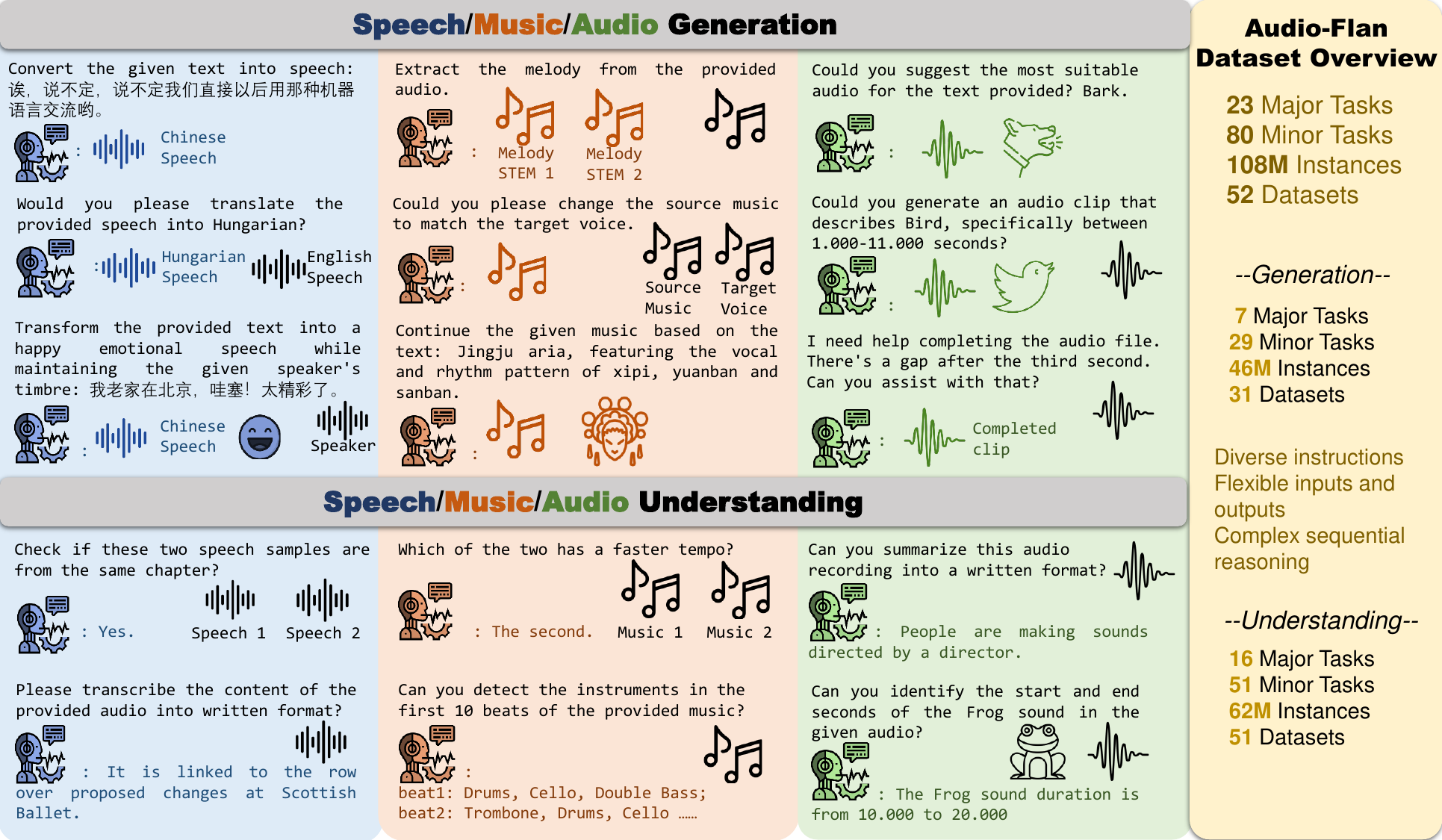}
}
\caption{Overview of Audio-FLAN dataset.}
\label{fig:dataset_overview}
\end{figure*}

Figure~\ref{fig:dataset_overview} illustrates the structure of the \textbf{Audio-FLAN} dataset, which spans a diverse range of tasks and instances. It is organized into 23 major tasks and 80 minor tasks from 52 released datasets\footnote{Each dataset may correspond to one or more tasks. The 52 datasets in Audio-FLAN represent the unique datasets after deduplication. The total number of data points for different tasks can exceed 52.}, totaling 108.5M instances. These tasks are divided into two primary categories: \textbf{understanding} and \textbf{generation}.

\begin{itemize}
    \item \textbf{Understanding}: This category consists of 16 major tasks and 51 minor tasks with 51 open-sourced datasets, amounting to 62.44M instances. The understanding tasks are further divided into three domains:
    \begin{itemize}
        \item \textbf{Speech}: 6 major tasks and 20 minor tasks, with 24 datasets and 57.42M instances.
        \item \textbf{Music}: 5 major tasks and 21 minor tasks, with 19 datasets and 1.46M instances.
        \item \textbf{Audio}: 5 major tasks and 10 minor tasks, with 8 datasets and 3.56M instances.
    \end{itemize}
    
    \item \textbf{Generation}: This category includes 7 major tasks and 29 minor tasks with 31 publicly available datasets, with a total of 46.06M instances. The generation tasks are categorized as follows:
    \begin{itemize}
        \item \textbf{Speech}: 2 major tasks, 14 minor tasks, with 12 datasets and 43M instances.
        \item \textbf{Music}: 2 major tasks, 7 minor tasks, with 13 datasets and 0.71M instances.
        \item \textbf{Audio}: 3 major tasks, 8 minor tasks, with 6 datasets and 2.35M instances.
    \end{itemize}
\end{itemize}

Overall, the \textbf{Audio-FLAN} dataset provides a comprehensive and balanced set of tasks across the \textbf{speech}, \textbf{music}, and \textbf{audio} domains, supporting both understanding and generation tasks in the audio field. The \textbf{Audio-FLAN} dataset fills a critical gap in the audio research community, offering the first large-scale, instruction-driven corpus for unified audio-language models. 

\subsection{Statistics of Task}
The \textbf{Audio-FLAN} dataset, spanning across the \textbf{speech}, \textbf{music}, and \textbf{audio} domains, is summarized in Table~\ref{tab:task_and_instances}. The dataset consists of 23 major tasks and 80 minor tasks across these domains, totaling 108.5M instances. These tasks cover a wide range of applications and modalities, integrating both \textbf{understanding} and \textbf{generation} tasks across various domains. The dataset’s diversity is further enhanced by the variety of input-output formats, including audio, text, and multimodal combinations such as audio and text, allowing it to represent complex and realistic scenarios.

\begin{table*}[h] 
  \centering  
  \resizebox{\textwidth}{!}{
  \begin{tabular}{ccccccc}  
    \toprule
     \textbf{Domain} &  \textbf{Major Task} & \#  \textbf{Minor Task} & \#  \textbf{Instances} &  \textbf{Input/Output} &  \textbf{U/G} \\  
    \midrule
    \multirow{8}{*}{Speech} & Speech Recognition    & 3 &  12.05M & audio/text & U \\  %
                            & Spoken Language Understanding     & 2 &  26.25M & audio/text & U \\  %
                            & Paralinguistic Attribute Recognition & 7 & 16.47M & audio/text & U \\
                            & Speaker Recognition   & 4 &  0.73M & audio/text & U \\
                            & Speech Caption    & 1 &  0.35M & audio/text & U \\
                            & Speech Detection  & 3 &  1.57M & audio/text & U \\
                            & Speech Enhancement    & 5 & 1.48M & audio/audio & G \\
                            & Speech Generation     & 9 &  41.52M & (audio, text)/audio & G \\  %
    \cline{2-6}
    \textbf{Total } & 8 & \textbf{34} & \textbf{100.42M} & - & - \\
    
    \midrule
    \multirow{7}{*}{Music} & Global MIR & 10 &  0.34M & audio/text & U \\  
                         & Sequential MIR & 3 & 0.43M & audio/text & U \\  
                         & Single Music Reasoning & 2 &  95.86K & audio/text & U \\  
                         & Multiple Music Reasoning & 5 & 0.57M & audio/text & U \\
                         & Music Caption & 1 & 28.21K & audio/text & U \\
                         & Music Separation & 2 & 40.26K & audio/audio & G \\
                         & Music Generation & 5 & 0.67M & (audio, text)/audio & G \\
    \cline{2-6}
    \textbf{Total}   & 7 & \textbf{28} & \textbf{2.17M} & - & - \\
    \midrule
    \multirow{8}{*}{Audio} & Audio Event Recognition & 4 &  1.30M & audio/text & U \\
                             & Audio Caption & 1 &  0.82M & audio/text & U \\
                             & Audio Advanced Understanding & 1     &  10K   & audio/text & U \\
                             & Audio Detection & 2      &  1.08M & audio/text & U \\
                             & Audio Classification & 2 &  0.38M & audio/text & U \\
                             & Audio Enhancement & 2    &  0.15M & audio/audio & G \\
                             & Audio Separation & 3     &  0.89M & audio/audio & G \\
                             & Audio Generation & 3     &  1.31M & (audio, text)/audio & G \\
    \cline{2-6}
    \textbf{Total} & 8  & \textbf{18} & \textbf{5.91M} & - & - \\
    \midrule
    \midrule
     \textbf{Total} & \textbf{23} & \textbf{80} &  \textbf{108.5M} &  - & - \\
    \bottomrule
  \end{tabular}
  }
  \caption{Detailed information of tasks and instances in Audio-FLAN. "U/G" indicates whether the task is for understanding (U) or generation (G). If the output is audio, it is classified as generation; otherwise, it is understanding.}  
  \vspace{-15pt}
  \label{tab:task_and_instances}  
\end{table*}

\textbf{Speech Domain:} The Speech domain encompasses 8 major tasks, including \textit{Speech Recognition}, \textit{Speech Generation}, and \textit{Paralinguistic Attribute Recognition}, addressing both understanding and generation tasks. The Speech domain includes 34 minor tasks, with a total of 100.42M instances, showcasing a comprehensive and diverse task representation. Notably, tasks such as \textit{Speech Enhancement} and \textit{Speech Generation} focus on generation tasks, while tasks like \textit{Speech Recognition} and \textit{Speaker Recognition} are geared toward understanding tasks. The large number of instances in this domain provides a rich dataset for training models, enhancing their ability to generalize across a wide range of speech-related tasks. This abundance of data enables models to learn robust representations, improving their performance and versatility when tackling unseen tasks in the Speech domain.

\textbf{Music Domain:} The Music domain features 7 major tasks, covering various music-related applications such as \textit{Global MIR} (Music Information Retrieval), \textit{Music Generation}, and \textit{Text-guided Music Generation}. Both understanding tasks (e.g., genre classification, emotion recognition) and generative tasks (e.g., music composition from text descriptions) are included. With 28 minor tasks and over 2.17 million instances, the Music domain excels in multi-modal tasks, such as \textit{Text-guided Music Generation}, where input combinations of text descriptions and audio prompts are used. The inclusion of music generation tasks involving multimodal inputs enhances the flexibility and capability of the unified model to generate and comprehend music in diverse ways. The variety in input-output combinations fosters a more comprehensive understanding of music, making the model highly adaptable and capable of handling both music-related understanding and generation tasks seamlessly.

\textbf{Audio Domain:} The Audio domain includes 8 major tasks, such as \textit{Audio Event Recognition}, \textit{Audio Generation}, and \textit{Audio Separation}, along with 18 minor tasks and 5.91 million instances. The tasks span a broad range of applications, from sound classification to audio enhancement and separation. Notably, the Audio domain includes tasks such as \textit{Audio Generation} and \textit{Audio Super-resolution}, which play a key role in advancing the field of audio processing. The diversity of tasks in this domain enhances the model’s ability to understand and generate a wide variety of audio content, further enriching the overall capabilities of the unified audio-language model.

The \textbf{Audio-FLAN} dataset makes a significant contribution to the development of unified models that can both understand and generate audio across multiple domains, including speech, music, and audio. By integrating a diverse set of tasks, the dataset ensures that the models can handle a broad spectrum of real-world audio applications. The varying number of instances across different tasks in the dataset provides a rich foundation for training models. Tasks with larger datasets, such as those in the speech domain, provide ample data for the model to develop a robust understanding of common patterns and features. This helps models generalize well across various tasks, improving their performance and robustness in real-world applications. The variety in instance sizes ensures that the model can remain adaptable and flexible, capable of learning from both high- and low-representation tasks, which is crucial for tasks that are less represented.

While the dataset is highly diverse, it is worth noting that the data distribution across domains is not perfectly balanced. The speech domain, with its larger number of instances, naturally provides more data for training compared to the music and audio domains. We are committed to continuously updating and expanding the \textbf{Audio-FLAN} dataset to include more tasks, domains, and instances. We also encourage the community to contribute by adding new tasks and improving the dataset. By working together, we can build a more comprehensive resource that further advances the development of unified audio-language models and benefits the broader research community.

\subsection{Distribution of Audio Attributes}

\begin{figure}[h] 
    \centering 
    \includegraphics[width=\linewidth]{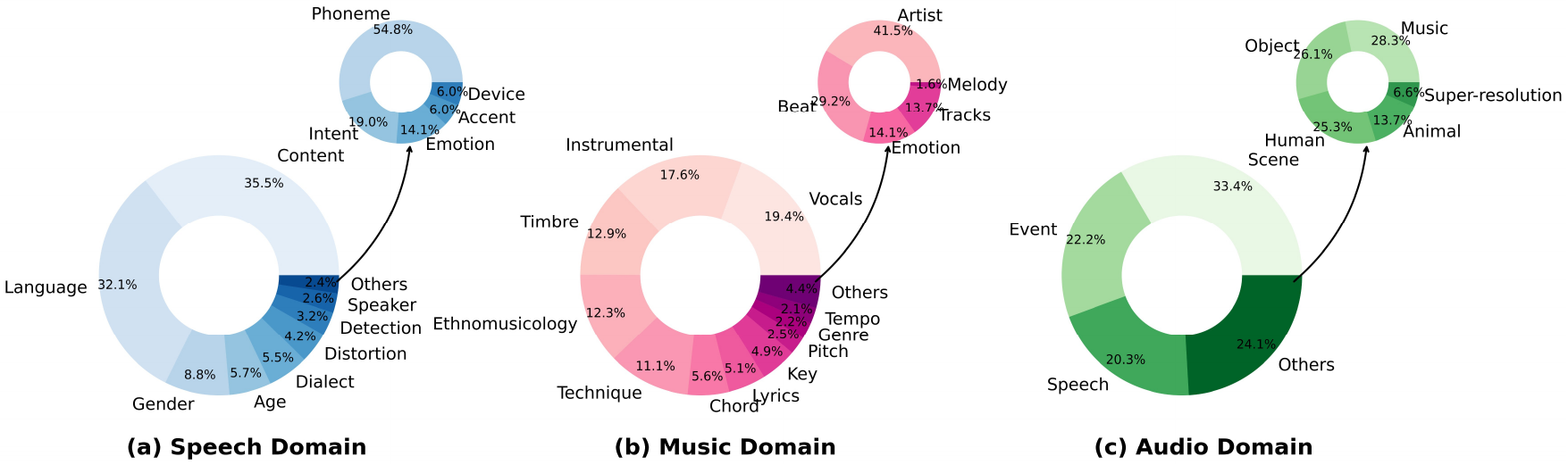} 
    \caption{Distribution of audio attributes in (a) speech domain, (b) music domain, and (c) audio domain.}
    \label{fig:data_type_distribution}
\end{figure}

 Each subdomain in the audio field encompasses a wide range of attributes. Specifically, the \textbf{speech} domain captures semantic content, speaker identity, and critical paralinguistic features such as emotion, language, accent, age, and more. The \textbf{music} domain contains a variety of musical attributes, including different instruments, timbres, techniques, and structures. Meanwhile, the \textbf{audio} domain covers diverse sounds, including events, animals, scenes, and even speech or music. To explore the different audio attributes in the \textbf{Audio-FLAN} dataset, we analyze the instance distribution of tasks related to these attributes across the speech, music, and audio domains, as shown in Figure~\ref{fig:data_type_distribution}.

\textbf{Speech Domain:}
As shown in Figure~\ref{fig:data_type_distribution} (a), in the speech domain, the most prominent features are \textbf{content} (35.5\%) and \textbf{language} (32.1\%). \textbf{content}-related tasks, like \textit{Automatic Speech Recognition (ASR)}, focus on transcribing spoken language into text, while \textbf{language}-related tasks, such as \textit{Language Identification} and \textit{Speech to Text Translation}, handle the translation and identification of speech across languages.

Additional tasks in the speech domain cover features like \textbf{gender} (8.8\%), which identifies the speaker’s gender, and \textbf{age} (5.7\%), \textbf{dialect} (5.5\%), and \textit{distortion} (4.2\%) tasks, such as \textit{Denoising} and \textit{Dereverberation}, which improve speech quality. Smaller, yet significant contributions come from tasks related to \textbf{emotion}, \textbf{accent}, and \textbf{device} (1.1\%), contributing to a more nuanced understanding of speech signals.

\textbf{Music Domain:}
As shown in Figure~\ref{fig:data_type_distribution} (b), the music domain's most prominent features are \textbf{instrumental} (17.6\%) and \textbf{timbre} (12.9\%). \textbf{instrumental} tasks, like \textit{Instrument Classification} and \textit{Beat-level Instrument Recognition}, focus on identifying and analyzing different musical instruments. \textbf{timbre} is related to the tonal quality of sound, and tasks like \textit{Singing Voice Conversion} capture the unique characteristics of sound sources.

The domain also includes \textbf{ethnomusicology} (12.3\%), which helps the model understand diverse cultural music, and tasks like \textit{Text-to-Music Generation} and \textit{Text-guided Music Continuation}. \textbf{vocals} (19.4\%) and \textbf{melody} (5.3\%) tasks like \textit{Vocal Technique Classification} and \textit{Melody Extraction} focus on analyzing vocal and melodic elements in music. Additional tasks cover \textbf{pitch} (5.1\%), \textbf{key} (4.9\%), and \textbf{chord} (2.2\%), focusing on musical structure and harmony.

\textbf{Audio Domain:}
As shown in Figure~\ref{fig:data_type_distribution} (c), the audio domain is dominated by \textbf{scene} (33.4\%), which represents environmental sounds, aiding in contextualizing audio. Tasks like \textit{Acoustic Scene Classification} categorize different environments based on their audio characteristics. \textbf{event} (22.2\%) and \textbf{speech} (20.3\%) features involve tasks like \textit{Sound Event Recognition} and \textit{Speech Detection}, which identify specific events and speech elements in general soundscapes.

Additionally, the \textbf{others} category (24.1\%) includes \textbf{music} (28.3\%), \textbf{object} (26.1\%), and \textbf{human} (25.3\%) features, covering tasks like \textit{Audio Event Detection}, \textit{Audio Source Separation}, and \textit{Speech and Non-speech Detection}, providing a comprehensive approach to general audio processing and recognition.

It is important to note that each instance may contain multiple features. As a result, the statistics presented reflect the frequency of feature occurrences rather than the absolute count of instances associated with each feature. 
This distribution highlights the \textbf{ rich diversity of attributes} within both the \textbf{speech}, \textbf{music} and \textbf{audio} domains, encompassing foundational tasks such as speech recognition and speaker identification, as well as more specialized areas like noise reduction, environmental sound recognition, and music analysis. The broad range of features and tasks in these domains supports the development of unified models that can be generalized across various audio-language tasks. This diversity enables models to adapt to a wide variety of contexts, enhancing their zero-shot generalization capabilities across different types of audio with diverse attributes.

\section{Conclusion and Discussion}


 The \textbf{Audio-FLAN} dataset represents a groundbreaking contribution to the audio domain by enabling \textit{instruction-tuning} for both \textbf{understanding} and \textbf{generation} tasks across the \textbf{speech}, \textbf{music}, and \textbf{audio} domains. This pioneering dataset consists of 23 major tasks and 80 minor tasks, with 16 major tasks dedicated to understanding and 7 major tasks focused on generation, totaling 108.5 million instances. By covering a wide array of tasks from speech recognition and emotion detection to music generation and audio event recognition, the \textbf{Audio-FLAN} dataset provides a comprehensive foundation for developing unified models that can handle both understanding and generation across multiple audio domains. This dataset is designed to support instruction-tuning, empowering models to follow complex audio instructions with minimal task-specific data. It paves the way for zero-shot generalization, enabling models to perform well on unseen tasks within and across domains, much like the advancements seen in text and vision models.


The \textbf{Audio-FLAN} dataset, while a major step towards unifying understanding and generation tasks across the speech, music, and audio domains, exhibits an imbalance in instance distribution. Understanding tasks, particularly in the speech domain, dominate the dataset, benefiting from well-established datasets and easier labeling. In contrast, generation tasks, such as text-to-audio or music generation, are more complex and less represented. This imbalance results in a greater number of instances in the speech domain, while the music and audio domains have fewer. This skew may lead to models being biased toward understanding tasks, potentially impacting their generalization to generation tasks or underrepresented domains.

Future work should focus on balancing the distribution of tasks across domains, ensuring a more even representation between \textit{understanding} and \textit{generation} tasks, especially in the music and audio domains. Additionally, expanding the dataset to include more tasks and incorporating additional datasets will strengthen the audio domain’s instruction-tuning capabilities, enhancing the development of unified models that can handle both understanding and generation tasks with improved zero-shot performance. Furthermore, integrating conversational data will be crucial for equipping models with the ability to engage in dynamic, real-time dialogue, broadening the dataset’s applicability to intelligent virtual agents and multimodal interaction systems.


\bibliographystyle{plainnat} 
\bibliography{main}

\appendix
\newpage
\onecolumn
\section{Appendix}
\label{sec:appendix}



\subsection{Task Definition}\label{appendix:task_definition}
\begin{center}
\textbf{Speech Domain}
\end{center}

Here, we provide a detailed list of each minor task definition for the speech, music, and audio domains, respectively.

\textbf{Speech Recognition (3 minor tasks)}

1. \underline{Automatic Speech Recognition}: transcribing speech into text.

2. \underline{Dialect Automatic Speech Recognition}: \underline{Automatic Speech Recognition} adapted for dialectal variations.


3. \underline{Phonetic Recognition}: identifying and classifying the smallest units of sound in spoken language, known as phonemes.




\textbf{Spoken Language Understanding (2 minor tasks)}

1. \underline{Intent Classification}: determining the purpose behind a user's spoken input.

2. \underline{Speech to Text Translation}: translating spoken language into written text in a different language.

\textbf{Paralinguistic Attribute Recognition (7 minor tasks)}

1. \underline{Gender Recognition}: classifying the biological gender of a speaker based on acoustic features of their voice. This task leverages acoustic features of speech, such as pitch, formant frequencies, and speech patterns, which tend to differ between male and female speakers due to physiological differences in the vocal tract and larynx.

2. \underline{Age Prediction}: estimating the age of a speaker based on the acoustic properties of their voice. This task utilizes various speech features, such as pitch, speaking rate, formant frequencies, and spectral characteristics, which can provide cues about the speaker's age. 

3. \underline{Emotion Recognition}: identifying and classifying the emotional state of a speaker based on their vocal expressions. 

4. \underline{Accent Recognition}: identifying the regional or cultural accent of a speaker based on their speech characteristics.

5. \underline{Spoken Paragraph Recognition}: determining whether two audio recordings contain the same spoken paragraph by analyzing the linguistic content.

6. \underline{Language Identification}: determining the language spoken from a given audio sample. 

7. \underline{Dialect Identification}: determining the specific dialect or regional variation of a language spoken in a given audio sample.

\textbf{Speaker Recognition (4 minor tasks)}

1. \underline{Speaker Verification}: verifying a speaker's identity by comparing their voice to a pre-recorded voiceprint (voice model) of the claimed identity. This process is used to authenticate or verify a speaker's identity, ensuring that the person speaking is who they claim to be. It includes text-independent and text-dependent speaker verification.

2. \underline{Speaker Diarization}: identifying "who spoke when" in an audio recording containing multiple speakers. This task segments an audio stream into homogeneous regions according to the speaker identity, effectively attributing each segment of speech to its corresponding speaker. 

3. \underline{Speaker Extraction}: extracting the speech of a target speaker from a mixture of sounds that may include multiple speakers and background noise.

4. \underline{Speaker Identification}: identifying a speaker from a set of known speakers based on their voice.

\textbf{Speech Caption (1 minor task)} 

1. \underline{Speech Caption}: generating synchronized text captions from spoken language.

\textbf{Speech Detection (3 minor tasks)}

1. \underline{Deepfake Detection}: detecting whether an audio clip has been artificially manipulated or synthesized using AI techniques, such as voice cloning or deepfake speech generation.

2. \underline{Vocoder Type Classification}: identifying and categorizing the type of vocoder used in a given speech signal.

3. \underline{Vocoder Type Classification}: identifying the device used to record a given speech segment based on its acoustic features.

\textbf{Speech Enhancement (5 minor tasks)}	

1. \underline{Denoising}: removing unwanted noise from an audio signal to enhance the clarity and quality of the speech. This task involves distinguishing between the speech signal and the background noise, which can include sounds like traffic, machinery, conversations, or other environmental noises. 

2. \underline{Dereverberation}: reducing or eliminating the effects of reverberation from an audio signal. Reverberation occurs when sound waves reflect off surfaces such as walls, ceilings, and floors, causing the original speech signal to be combined with multiple delayed copies of itself. 

3. \underline{Declipping}: restoring audio signals that have been distorted due to clipping. Clipping occurs when the amplitude of an audio signal exceeds the maximum limit that a recording or playback system can handle, causing the peaks of the waveform to be "clipped" off.

4. \underline{Speech Bandwidth Extension}: enhancing narrowband speech quality by extending its frequency range. Narrowband speech often lacks the higher frequencies that contribute to the naturalness and clarity of speech.

5. \underline{Signal-to-noise Ratio Estimation}: quantifying the ratio of the power of a signal to the power of background noise. This task provides a quantitative measure of the quality of a signal.

\textbf{Speech Generation (9 minor tasks)}

1. \underline{Text to Speech}: converting written text into spoken words. It involves synthesizing speech that is natural and understandable, enabling computers to "read" text aloud.

2. \underline{Zero-shot Text to Speech/Voice Cloning}: generating synthetic speech for voices or styles it has never encountered during training. 

3. \underline{Emotional Text to Speech}: synthesizing speech with emotional nuances. The goal is to produce speech that not only conveys the content of the text but also expresses specific emotions, making the synthetic voice more engaging and human-like.
		
4. \underline{Zero-shot Emotional Text to Speech}: generating emotional speech that adapts to an unseen speaker’s voice while rendering specified emotions. 

5. \underline{Descriptive Speech Synthesis}: generating synthetic speech that not only replicates the spoken content but also conveys descriptive information about the context of the speech, such as emotions, tone, or other paralinguistic features.

6. \underline{Spontaneous Text to Speech}: generating synthetic speech that mimics the characteristics of spontaneous unscripted human speech. Spontaneous TTS aims to replicate the naturalness, variability, and informal aspects of everyday conversational speech. This includes features such as hesitations, fillers (e.g., "um," "uh"), varying speech rates, and natural prosody changes.

7. \underline{Voice Conversion}: converting one speaker's voice to resemble another's while preserving linguistic content and prosody.

8. \underline{Emotion Conversion}: transforming the emotional tone of a spoken utterance from one emotion to another while preserving the linguistic content. 

9. \underline{Speech to Speech Translation}: converting spoken language in one language directly into spoken language in another language.

\begin{center}
\textbf{Music Domain}
\end{center}

\textbf{Global MIR (10 minor tasks)}: 

1. \underline{Key Detection}: recognizing the key signature of the given music.

2. \underline{Scale Recognition}: recognizing the scale of the given music.

3. \underline{Music Tagging}: assigning descriptive tags to audio files, such as genre, style, tempo, key, artist, and emotion.

4. \underline{Genre Classification}: categorizing the music into certain genres.

5. \underline{Emotion Classification}: recognizing emotion categories from the music.

6. \underline{Pitch Classification}: classifying the pitch of the given audio.

7. \underline{Instrument Classification}: identifying all existing instruments from the music.

8. \underline{Vocal Technique Classification}: detecting the playing techniques used in the vocal music.

9. \underline{Instrumental Technique Classification}: detecting the playing techniques used in the instrumental music.

10 \underline{Artist Identification}: identifying the relevant artists of a piece of music, given a set of artists as the options.

\textbf{Sequential MIR (3 minor tasks)}

1. \underline{Beat Tracking}: detecting and aligning beats of a music excerpt.

2. \underline{Chord Estimation}: estimating the chords sequence at each time step of a music excerpt.

3. \underline{Progression Extraction}: extracting the chord progression represented by chord number sequence. 

\textbf{Single Music Reasoning (2 minor tasks)}	

1. \underline{Beat-level Instruments Recognition}: recognizing the instruments from a certain beat or a certain segment.


2. \underline{Beat-level Pitch Estimation}: estimating the pitch of a certain beat or segment.

\textbf{Multiple Music Reasoning (5 minor tasks)}

1. \underline{Tempo Comparison}: comparing the tempo characteristics between two music excerpts.

2. \underline{Instruments Comparison}: comparing instruments of two music excerpts.

3. \underline{Key Comparison}: comparing keys of two music excerpts.

4. \underline{Instrumental Technique Comparison}: comparing playing techniques of two music excerpts.

5. \underline{Emotion Comparison}: comparing emotions of two excerpts.

\textbf{Music Caption (1 minor task)}

1. \underline{Music Caption}: generating textual descriptions for a piece of music.

\textbf{Music Separation (2 minor tasks)}

1. \underline{Melody Extraction}: extracting the melody at each time step from a music excerpt.

2. \underline{Text-guided Source Separation}: separate certain tracks from a piece of mixed music with the text instruction.

\textbf{Music Generation (5 minor tasks)}


1. \underline{Text-to-Music Generation}: generating the music given the text caption.
 



2. \underline{Text-guided Music Continuation}: extending a given initial audio segment based on a textual description of musical characteristics while ensuring continuity and coherence.
		
3. \underline{Lyrics-to-song Generation}: composing a song with the vocal track and instrumental track based on the given lyrics.

4. \underline{Singing Voice Synthesis}: synthesizing the voice given the pitches and lyrics sequence.

5. \underline{Singing Voice Conversion}: transforming the vocals (including the lyrics and melody) of singer A(source vocals) to sound like Singer B (target singer).

\vspace{10pt}
\begin{center}
\textbf{Audio Domain}
\end{center}

\textbf{Audio Event Recognition (4 minor tasks)}

1. \underline{Sound Event Sequence Recognition}: identifying and sequencing various sounds in an audio stream.

2. \underline{Sound Event Recognition}: detecting and identifying a particular sound in audio data.

3. \underline{Sound Event Detection}: determining when a specific sound occurs within an audio clip.

4. \underline{Acoustic Scene Classification}: classifying audio by its acoustic environment (e.g., park, street).




\textbf{Audio Caption (1 minor task)}

1. \underline{Audio Caption}: generating text descriptions that summarize or explain the content of an audio clip.



\textbf{Audio Advanced Understanding (1 minor task)}


1. \underline{Sound Event Understanding}: extracting meaningful information from multiple audio signals (e.g. What is happening in the given audio). 


\textbf{Audio Detection (2 minor tasks)}

1. \underline{Deepfake Audio Detection}: identifying synthetic or manipulated audio content.


2. \underline{Voice Activity Detection}: identifying segments where human speech is present in the given audio.

\textbf{Audio Classification (2 minor tasks)}

1. \underline{Speech, Silence, Music and Noise Classification}: distinguishing between music, speech, and various types of noise.

2. \underline{Speech and Non-speech Detection}: identifying segments which contain speech or non-speech of the given audio.

\textbf{Audio Enhancement (2 minor tasks)}	

1. \underline{Audio Inpainting}: filling in missing parts of an audio signal.

2. \underline{Audio Super-resolution}: improving the perceptual quality of an audio signal by increasing its resolution.

\textbf{Audio Separation (3 minor tasks)}

1. \underline{Text-guided Audio Source Separation}: isolating target sounds from audio using text prompts.

2. \underline{Label-querying Sound Extraction}: extracting sounds belonging to a predefined category from an audio mixture, given a textual label 

3. \underline{Audio-querying Sound Extraction}: separating target sounds using an audio reference.

\textbf{Audio Generation (3 minor tasks)}	

1. \underline{Text-guided Audio Generation}: creating audio based on a textual description.

2. \underline{Time-grounded Text-to-audio Generation}: generating time-aligned audio from text prompts.

3. \underline{Audio Continuation}: generating content that smoothly extends a given audio clip.

\begingroup
\let\clearpage\relax
\subsection{Datasets for Each Task}
Here, we present the datasets associated with each minor task.
\onecolumn
\begin{longtable}{p{1cm}p{6cm}p{6cm}}   
\caption{Minor task and its corresponding datasets}  
\label{tab:task2dataset}  
\\  
\hline
\textbf{Domain} &  \textbf{Minor Task} &   \textbf{Dataset} \\
\hline
\endfirsthead  
\multicolumn{3}{|c|}{\textit{Continued from the previous page}} \\  
\hline
\textbf{Domain} &  \textbf{Minor Task}  &  \textbf{Dataset} \\
\hline
\endhead  
\hline
\endfoot  
\hline
\endlastfoot  

Speech & Automatic Speech Recognition & Aishell1~\citep{bu2017aishell}, Aishell2~\citep{du2018aishell}, Aishell3~\citep{shi2020aishell}, ESD~\citep{zhou2022emotional}, EmoV\_DB~\citep{adigwe2018emotional}, FLEURS~\citep{conneau2023fleurs}, Fluent Speech Commands~\citep{lugosch2019speech}, HQ-Conversations~\citep{xia2024iscslp}, HiFi TTS~\citep{bakhturina2021hi}, LJSpeech~\citep{ljspeech17}, MLS~\citep{Pratap2020MLSAL}, The Parallel Audiobook Corpus~\citep{pacorpus18}, VCTK~\citep{veaux2017cstr}, aidatatang~\citep{aidatatang200zh}, common voice~\citep{ardila2019common}, LibriTTS-R~\citep{koizumi2023librittsr} \\
    \cline{2-3} %

    \multirow{35}{*}{} & Dialect Automatic Speech Recognition & KeSpeech~\citep{tang2021kespeech} \\  \cline{2-3}


    & Phonetic Recognition & Aishell3~\citep{shi2020aishell}, LibriTTS-R~\citep{koizumi2023librittsr} \\ \cline{2-3}

    & Intent Classification &  Fluent Speech Commands~\citep{qian2021speech} \\ \cline{2-3} 



    & Gender Recognition & Aishell1~\citep{bu2017aishell}~\citep{bu2017aishell}, Aishell2~\citep{du2018aishell}, Aishell3~\citep{shi2020aishell}, Fluent Speech Commands~\citep{lugosch2019speech}, HQ-Conversations~\citep{xia2024iscslp}, KeSpeech~\citep{tang2021kespeech}, The Parallel Audiobook Corpus~\citep{pacorpus18}, LibriTTS-R~\citep{koizumi2023librittsr} \\
    \cline{2-3} %
    & Age Recognition & HQ-Conversations~\citep{xia2024iscslp}, KeSpeech~\citep{tang2021kespeech} \\
    \cline{2-3} %
    & Emotion Recognition & ESD~\citep{zhou2022emotional} \\
    \cline{2-3} %
    & Accent Recognition & HQ-Conversations~\citep{xia2024iscslp} \\
    \cline{2-3} %
    & Spoken Paragraph Recognition & LibriTTS-R~\citep{koizumi2023librittsr} \\
    \cline{2-3} %
    & Language Identification & Aishell1~\citep{bu2017aishell}~\citep{bu2017aishell}, Aishell2~\citep{du2018aishell}, Aishell3~\citep{shi2020aishell}, ESD~\citep{zhou2022emotional}, EmoV\_DB~\citep{adigwe2018emotional}, FLEURS~\citep{conneau2023fleurs}, HQ-Conversations~\citep{xia2024iscslp}, HiFi TTS~\citep{bakhturina2021hi}, LJSpeech~\citep{ljspeech17}, MLS~\citep{Pratap2020MLSAL}, The Parallel Audiobook Corpus~\citep{pacorpus18}, aidatatang~\citep{aidatatang200zh}, common voice~\citep{ardila2019common}, LibriTTS-R~\citep{koizumi2023librittsr} \\
    \cline{2-3} %
    & Dialect Identification & KeSpeech~\citep{tang2021kespeech} \\
    \cline{2-3} %
    & Speaker Verification & Aishell1~\citep{bu2017aishell}~\citep{bu2017aishell}, Aishell2~\citep{du2018aishell}, Aishell3~\citep{shi2020aishell}, ESD~\citep{zhou2022emotional}, EmoV\_DB~\citep{adigwe2018emotional}, Fluent Speech Commands~\citep{lugosch2019speech}, HQ-Conversations~\citep{xia2024iscslp}, HiFi TTS~\citep{bakhturina2021hi}, KeSpeech~\citep{tang2021kespeech}, The Parallel Audiobook Corpus~\citep{pacorpus18}, LibriTTS-R~\citep{koizumi2023librittsr} \\
    \cline{2-3} %
    & Speaker Diarization & AliMeeting~\citep{Yu2022M2MeT} \\
    \cline{2-3} %
    & Speaker Extraction & LibriMix~\citep{cosentino2020librimix}\\
    \cline{2-3} %
    & Speaker Identification & KeSpeech~\citep{tang2021kespeech} \\
    \cline{2-3} %
    & Speech Caption & LibriTTS-R~\citep{koizumi2023librittsr} \\
    \cline{2-3} %
    & Deepfake Detection & ASVSpoof2021~\citep{liu2023asvspoof} \\
    \cline{2-3} %
    & Vocoder Type Classification & ASVSpoof2021~\citep{liu2023asvspoof} \\  \cline{2-3} %
    & Device Recognition &  HQ-Conversations~\citep{xia2024iscslp} \\  \cline{2-3} %

    & Denoising & DNS~\citep{reddy2001interspeech} \\
    \cline{2-3} %
    & Dereverberation & DNS~\citep{reddy2001interspeech} \\
    \cline{2-3} %
    & Declipping & DNS~\citep{reddy2001interspeech} \\
    \cline{2-3} %
    & Speech Bandwidth Extension & DNS ~\citep{reddy2001interspeech} \\
    \cline{2-3} %
    & Signal-to-noise Ratio Estimation & LibriTTS-R~\citep{koizumi2023librittsr} \\
    \cline{2-3} %

    & Speech to Text Translation & CVSS~\citep{jia2022cvss}, FLEURS~\citep{conneau2023fleurs} \\
    \cline{2-3} %
    & Text to Speech & Aishell1~\citep{bu2017aishell}~\citep{bu2017aishell}, Aishell2~\citep{du2018aishell}, Aishell3~\citep{shi2020aishell}, ESD~\citep{zhou2022emotional}, EmoV\_DB~\citep{adigwe2018emotional}, FLEURS~\citep{conneau2023fleurs}, Fluent Speech Commands~\citep{lugosch2019speech}, HQ-Conversations~\citep{xia2024iscslp}, HiFi TTS~\citep{bakhturina2021hi}, KeSpeech~\citep{tang2021kespeech}, LJSpeech~\citep{ljspeech17}, MLS~\citep{Pratap2020MLSAL}, The Parallel Audiobook Corpus~\citep{pacorpus18}, VCTK~\citep{veaux2017cstr}, aidatatang~\citep{aidatatang200zh}, common voice~\citep{ardila2019common}, LibriTTS-R~\citep{koizumi2023librittsr}, Genshin~\citep{GenshinDatasets}, StarRail~\citep{StarRailDatasets} \\ 
    \cline{2-3}  
    & Zero-shot Text to Speech & Fluent Speech Commands~\citep{lugosch2019speech}, LibriTTS-R~\citep{koizumi2023librittsr},HQ-Conversations~\citep{xia2024iscslp},Fluent Speech Commands~\citep{lugosch2019speech}, Aishell2~\citep{du2018aishell}, Aishell3~\citep{shi2020aishell}, KeSpeech~\citep{tang2021kespeech}  \\
    \cline{2-3} %
    & Emotional Text to Speech & ESD~\citep{zhou2022emotional}, EmoV\_DB~\citep{adigwe2018emotional}  \\
    \cline{2-3} %
    & Zero-shot Emotional Text to Speech & ESD~\citep{zhou2022emotional} \\
    \cline{2-3} %
    & Descriptive Speech Synthesis & LibriTTS-R~\citep{koizumi2023librittsr} \\
    \cline{2-3} %
    & Voice Conversion & ESD~\citep{zhou2022emotional} \\
    \cline{2-3} %
    & Emotion Conversion &  ESD~\citep{zhou2022emotional} \\
    \cline{2-3} %
    & Speech to Speech Translation & FLEURS~\citep{conneau2023fleurs} \\
    \cline{2-3} %
    
\midrule

Music & Key Detection & AAM~\citep{ostermann2023aam}, FreeSound Loop Dataset~\citep{ramires2020}\\
    \cline{2-3} %
    \multirow{29}{*}{} & Music Tagging & MTG~\citep{bogdanov2019mtg} \\
    \cline{2-3} %
    & Genre Classification & CSD~\citep{choi2020children}, MTG~\citep{bogdanov2019mtg},FreeSound Loop Dataset~\citep{ramires2020} \\
    \cline{2-3} %
    & Emotion Classification & MTG~\citep{bogdanov2019mtg} \\
    \cline{2-3} %
    & Pitch Classification & NSynth~\citep{engel2017neural} \\
    \cline{2-3} %
    & Instrument Classification & AAM~\citep{ostermann2023aam}, MTG~\citep{bogdanov2019mtg}, NSynth~\citep{engel2017neural} \\
    \cline{2-3} %
    & Vocal Technique Classification & VocalSet~\citep{wilkins2018vocalset} \\
    \cline{2-3} %
    & Instrumental Technique Classification & CCOM-HuQin~\citep{zhang2022ccom} \\
    \cline{2-3} %
    & Artist Identification & FMA~\citep{defferrard2016fma}\ \\
    \cline{2-3} %
    
    & Beat Tracking & AAM~\citep{ostermann2023aam} \\
    \cline{2-3} %
    & Melody Extraction & MedleyDB~\citep{bittner2014medleydb} \\
    \cline{2-3} %
    & Chord Estimation & AAM~\citep{ostermann2023aam} \\ \cline{2-3} 

    & Beat-level Instrument Recognition & AAM~\citep{ostermann2023aam} \\ \cline{2-3} %
    & Progression Extraction & JazzNet~\citep{adegbija2023jazznet}\\  \cline{2-3} %
    & Scale Recognition & JazzNet~\citep{adegbija2023jazznet}\\  \cline{2-3} %

    & Beat-level Pitch Estimation & AAM~\citep{ostermann2023aam}, CSD~\citep{choi2020children}, Vocadito~\citep{bittner2021vocadito} \\  \cline{2-3} %

    & Tempo Comparison & GTZAN Rhythm~\citep{marchand2015gtzan}, FreeSound Loop Dataset~\citep{ramires2020} \\  \cline{2-3} %

    & Instrument Comparison & NSynth~\citep{engel2017neural} \\  \cline{2-3} %

    & Key Comparison & GiantSteps Key~\citep{knees2015two} \\  \cline{2-3} %

    & Emotion Comparison & MTG~\citep{bogdanov2019mtg} \\  \cline{2-3} %

    & Instrumental Technique Comparison & CCOM-HuQin~\citep{zhang2022ccom} \\  \cline{2-3} %


    & Music Caption & Musiccaps~\citep{agostinelli2023musiclm}, FreeSound Loop Dataset~\citep{ramires2020} \\  \cline{2-3} %


    & Text-to-music Generation & FreeSound Loop Dataset~\citep{ramires2020}, Musiccaps~\citep{agostinelli2023musiclm}, Compmusic~\citep{comp_srinivasamurthy2021saraga, comp_anantapadmanabhan2013modal, comp_black2014automatic, comp_caro2018musical, comp_gupta2015discovery, comp_koduri2014intonation, comp_kuriakose2015akshara, comp_pretto2018nawba, comp_srinivasamurthy2014supervised, comp_srinivasamurthy2015particle, comp_srinivasamurthy2016generalized} \\  \cline{2-3} %




    & Text-guided Music Continuation &  Compmusic~\citep{comp_srinivasamurthy2021saraga, comp_anantapadmanabhan2013modal, comp_black2014automatic, comp_caro2018musical, comp_gupta2015discovery, comp_koduri2014intonation, comp_kuriakose2015akshara, comp_pretto2018nawba, comp_srinivasamurthy2014supervised, comp_srinivasamurthy2015particle, comp_srinivasamurthy2016generalized} \\  \cline{2-3} %

    & Lyrics2song Generation & CSD~\citep{choi2020children}, Vocadito~\citep{bittner2021vocadito}, Opencpop~\citep{wang2022opencpop}, Opensinger~\citep{huang2021multi} \\  \cline{2-3} %

    & Singing Voice Synthesis & CSD~\citep{choi2020children}, Vocadito~\citep{bittner2021vocadito}, Opencpop~\citep{wang2022opencpop}, Opensinger~\citep{huang2021multi} \\  \cline{2-3} %

    & Singing Voice Conversion & Opensinger~\citep{huang2021multi}, m4singer~\citep{zhang2022m4singer} \\  \cline{2-3} %

    & Text-guided Source Separation & MedleyVox~\citep{jeon2023medleyvox}, Moises~\citep{pereira2023moisesdb} \\ 
\midrule

Audio & Sound Event Sequence Recognition & Audioset~\citep{45857} \\ \cline{2-3} %

    \multirow{17}{*}{} & Acoustic Scene Classification & TAU Urban Acoustic Scenes~\citep{heittola2020acoustic} \\ \cline{2-3} %



    & Audio Caption &  Audioset~\citep{45857}, Freesound~\citep{font2013freesound} \\ \cline{2-3} %

    & Text-guided Audio Generation & Audioset~\citep{45857}, Freesound~\citep{font2013freesound} \\ \cline{2-3} %

    & Time-grounded Text-to-audio Generation & Audioset~\citep{45857} \\ \cline{2-3} %
    
    & Audio Continuation & Wavcaps~\citep{mei2023wavcaps} \\ \cline{2-3} %

    & Audio Inpainting & Audioset~\citep{45857} \\ \cline{2-3} %

    & Audio Super-resolution & Audioset~\citep{45857} \\ \cline{2-3} %

    & Sound Event Understanding & Vocal Imitation~\citep{kim2018vocal} \\ \cline{2-3} %

    & Text-guided Audio Source Separation & Wavcaps~\citep{mei2023wavcaps} \\ \cline{2-3} %

    & Label-querying Sound Extraction & VGG~\citep{chen2020vggsound} \\ \cline{2-3} %

    & Audio-querying Sound Extraction & VGG~\citep{chen2020vggsound}\\ \cline{2-3} %

    & Deepfake Audio Detection & ADD2023~\citep{yi2023add}\\ \cline{2-3} %

    & Voice Activity Detection & DNS for VAD~\citep{reddy2001interspeech} \\ \cline{2-3} %

    & Speech, Silence, Music and Noise Classification & Audioset~\citep{45857} \\ \cline{2-3} %

    & Speech Nonspeech Detection & Wavcaps~\citep{mei2023wavcaps}\\
\bottomrule

\end{longtable}

\onecolumn
\begin{longtable}{p{2cm}p{6cm}p{6cm}}   
\caption{Detailed information of datasets.}  
\label{tab:dataset_hours}  
\\  
\hline
\textbf{Domain} &  \textbf{Dataset} &   \textbf{Audio Length (\#hours)} \\
\hline
\endfirsthead  
\multicolumn{3}{|c|}{\textit{Continued from the previous page}} \\  
\hline
\textbf{Domain} &  \textbf{Dataset}  &  \textbf{Audio Length (\#hours)} \\
\hline
\endhead  
\hline
\endfoot  
\hline
\endlastfoot  

\multirow{22}{*}{Speech} & common voice \citep{ardila2019common}  & $19, 673$  \\  
                         & aidatatang \citep{aidatatang200zh}  & $200$ \\  
                         & libritts-R \citep{koizumi2023librittsr}  & $585$ \\  
                         & libritts \citep{zen2019libritts}  & $586$ \\  
                         & HQ-Conversations \citep{xia2024iscslp}  & $100$ \\  
                         & EmoV\_DB \citep{adigwe2018emotional}  & $9.49$ \\  
                         & VCTK \citep{veaux2017cstr}  & $44$ \\  
                         & MLS \citep{Pratap2020MLSAL}  & $45, 042$ \\  
                         & FLEURS \citep{conneau2023fleurs}  & $17$ \\  
                         & Fluent speech commands \citep{lugosch2019speech}  & $19$ \\  
                         & LibriMix \citep{cosentino2020librimix}  & $500$ \\  
                         & Aishell1 \citep{bu2017aishell}  & $155$ \\  
                         & Aishell2 \citep{du2018aishell}  & $1,036$  \\  
                         & Aishell3 \citep{shi2020aishell}  & $65$  \\  
                         & LJSpeech \citep{ljspeech17}  & $23.9$ \\  
                         & The Parallel Audiobook Corpus \citep{pacorpus18}  & $121$  \\  
                         & HiFi TTS \citep{bakhturina2021hi}  & $291.6$ \\  
                         & KeSpeech \citep{tang2021kespeech}  & $1, 428$ \\  
                         & ESD \citep{zhou2022emotional}  & $29$ \\  
                         & CVSS \citep{jia2022cvss}  & $3, 809$ \\  
                         & ASVSpoof2021 \citep{liu2023asvspoof}  & $1270.5$  \\    
    \midrule
    \multirow{19}{*}{Music} & Opencpop \citep{wang2022opencpop}  & $5.2$ \\  %
                           & m4singer \citep{zhang2022m4singer}  & $29.77$  \\  %
                           & FreeSound Loop Dataset \citep{ramires2020}  & $34.7$ \\  %
                           & Opensinger \citep{huang2021multi}  & $50$  \\  %
                           & MedleyVox \citep{jeon2023medleyvox}  & $1.1$ \\  %
                           & Vocadito \citep{bittner2021vocadito}  & $0.23$ \\  %
                           & MoisesDB \citep{pereira2023moisesdb}  & $14.4$ \\  %
                           & CSD \citep{choi2020children}  & $4.86$ \\  %
                           & Musiccaps \citep{agostinelli2023musiclm}  & $15.28$\\  %
                           
                           & GTZAN rhythm \citep{marchand2015gtzan}  & $8.3$ \\  %
                           & GiantSteps key \citep{knees2015two}  & $20.07$ \\  %
                           & CCOM-HuQin \citep{zhang2022ccom}  & $4.3$ \\  %
                           & NSynth \citep{engel2017neural}  & $340$ \\  %
                           & MedleyDB \citep{bittner2014medleydb}  & $7.45$ \\  %
                           & Free Music Archive \citep{fma_dataset}  & $8, 232$ \\  %
                           & AAM \citep{ostermann2023aam}  & $125$ \\  %
                           & MTG \citep{bogdanov2019mtg}  & $3,777$ \\  %

    \midrule
    \multirow{11}{*}{Audio}  & Audioset \citep{45857}  & $5208$ \\  %
                            & VGGSound \citep{chen2020vggsound}  & $550$ \\  %
                            & Wavcaps \citep{mei2023wavcaps}  & $3, 793.3$\\  %
                            & freesound \citep{font2013freesound}  & $6446.05$ \\  %
                            & TAU Urban Acoustic Scenes \citep{heittola2020acoustic}  & $68.18$ \\  %
                            & Vocal Imitation \citep{kim2018vocal}  & $24$ \\  %
                            & ESC \citep{piczak2015dataset}  & $2.78$ \\  %
                            & DNS for VAD \citep{reddy2001interspeech}  & $562.72$ \\  %
                            & ADD2023 \citep{yi2023add} &   $220$ \\

    \midrule
     \textbf{Total}& \textbf{50} &  \textbf{104,549.18} \\
    \bottomrule
    
\end{longtable}

\endgroup

\subsection{Instruction Template}
\label{appendix:instruction_template}
Here we provide the complete task instruction template in JSONL format with all fields. 


\begin{tcolorbox}[colback=white, colframe=black, boxrule=0.2mm, arc=0mm, title=Speech-to-Text Translation]
\begin{lstlisting}[basicstyle=\ttfamily, breaklines=true]
{"instruction": "Please translate the speech into the text in English.", 
  "input": "<|SOA|>Speech_Audio<|EOA|>", 
  "output": "Nevertheless, there are many distinctive ways of drinking coffee around the world that are worth experiencing.", 
  "uuid": "UUID", 
  "split": ["train"], 
  "task_type": {
    "major": ["Spoken Language Understanding"], 
    "minor": ["Speech-to-text Translation"], 
    "U/G": ["understanding"], 
    "unseen": false
  }, 
  "domain": "speech", 
  "source": ["unknown"]
"other": null}
\end{lstlisting}
\end{tcolorbox}

\begin{tcolorbox}[colback=white, colframe=black, boxrule=0.2mm, arc=0mm, title=Text-guided Music Continuation]
\begin{lstlisting}[basicstyle=\ttfamily, breaklines=true]
{"instruction": "Please continue the audio music prompt based on the given text description", 
"input": "This is a Carnatic music piece set in the atana raga. It follows the 5/8 meter and is composed in the khandaChapu taala. The lead instrument featured in this performance is vocal, accompanied by Mridangam. The kalai of this composition is 1.\n audio prompt: <|SOA|>Music_Audio<|EOA|>", 
"output": "audio: <|SOA|>Musi_Audio<|EOA|>", 
"uuid": "UUID", 
"split": ["test"], 
"task_type": {
    "major": ["Music Generation"], 
    "minor": ["Text-guided Music Continuation"], 
    "U/G": ["generation"], 
    "unseen": false
    }, 
"domain": "music", 
"source": ["unknown"], 
"other": null}
\end{lstlisting}
\end{tcolorbox}

\begin{tcolorbox}[colback=white, colframe=black, boxrule=0.2mm, arc=0mm, title=Sound Super-resolution]
\begin{lstlisting}[basicstyle=\ttfamily, breaklines=true]
{"instruction": "Please increase the resolution of the given audio signal to 32k Hz.", 
"input": "audio: <|SOA|>Sound_Audio<|EOA|>.", 
"output": "<|SOA|>Sound_Audio<|EOA|>", 
"uuid": "UUID", 
"split": ["train"], 
"task_type": {
    "major": ["Sound Generation"], 
    "minor": ["Sound Super-resolution"],
    "U/G": ["generation"], 
    "unseen": false
    }, 
"domain": "audio", 
"source": ["youtube"], 
"other": null}
\end{lstlisting}
\end{tcolorbox}

The definitions of each field are described as follows:

\textbf{Instruction}: this field provides the instructions for the task, outlining the specific operation to be performed.

\textbf{Input}: this field contains the input data for the task, which represents the raw information to be processed.

\textbf{Output}: this field represents the expected result or outcome after processing the input data.

\textbf{Uuid}: this field assigns a unique identifier to each task instance, enabling the system to track and manage individual tasks.

\textbf{Split}: this field specifies the dataset partition for the task, such as "train", "test", or "dev", which correspond to the training, testing, and development datasets, respectively.

\textbf{Task\_type}: this field outlines the nature of the task:

- \textbf{Major}: indicates the primary category of the task.

- \textbf{Minor}: specifies the secondary or more specific task.

- \textbf{U/G}: distinguishes whether the task focuses on generation or understanding.

- \textbf{Unseen}: a boolean value that indicates whether the task involves data that has not been encountered before.

\textbf{Domain}: this field defines the domain in which the task is situated, such as "speech", "music", or "audio".

\textbf{Source}: this field identifies the origin of the audio, such as "audiobook", "youtube", or "studio", signifying where the audio signal is sourced from.

\textbf{Other}: this field can store any additional metadata relevant to the task, if applicable.

\subsection{Instruction Variation Prompt}
\label{appendix:instruction_variation_prompt}
As mentioned in Section~\ref{sec:self_instruct}, all instances are automatically varied by entering a standard prompt in the existing LLMs, which is presented as follows. 


\begin{tcolorbox}[
    colframe=black,  
    colback=white,   
    boxrule=0.5mm,    
    arc=0mm,          
    sharp corners,     
    breakable
]

You are tasked with paraphrasing the values of the following fields: "instruction", "input", and "output". Your goal is to generate varied and creative rewrites for each of these fields. Please adhere to the following guidelines:

\begin{enumerate}
    \item \textbf{Paraphrase Instructions}:
    \begin{itemize}
        \item Paraphrase the "instruction" field in diverse ways by changing the sentence structure, style, and tone. Use a variety of sentence types, including:
        \begin{itemize}
            \item Direct commands (e.g., "Turn this into speech.")
            \item Polite requests (e.g., "Could you please convert this to speech?")
            \item Questions (e.g., "Can you turn this into audio?")
            \item Suggestions (e.g., "It would be great if you could convert this.")
            \item Exclamations or emphatic forms (e.g., "I really need this to be in audio form.")
        \end{itemize}
        \item Feel free to add polite elements, such as "please," "kindly," or "if you would be so kind," as long as they remain natural.
    \end{itemize}

    \item \textbf{Paraphrase Inputs}:
   
    \begin{itemize}
        \item  Change the labels for fields like "text:"`, "text\_description:"`, "audio:"`, "speaker\_audio:"`, "audio\_sample1"`, "audio\_sample2"` etc., according to "instruction", while retaining their original meaning. Examples include:
        \begin{itemize}
            \item "text:" to "spoken text," "speech input," "text excerpt," etc.
            \item "text\_description:" to "voice style," "descriptive text," "tone characteristics," etc.
            \item "audio:" to "source audio," "reference speech," "given recording," etc.
            \item "speaker\_audio:" to "speaker prompt," "reference voice," "voice sample," etc.
            \end{itemize}
        \item Ensure that the content following "text:" remains semantically identical to the original. The content following each label should remain unchanged, with only the labels varying.
    \end{itemize}

    \item \textbf{Maintain Consistency in Outputs}:
    \begin{itemize}
        \item Depending on the tone of the instruction, introduce additional phrases such as:
        \begin{itemize}
            \item "The gender is ", "Gender: ".
            \item "The language is ", "Language in the given speech is ".
            \item "The speakers in the given two speechs are ", "The anwser is ".
            \item "Transcription is: ", "The text of the given speech is: ".
            \item "IPA Phonemes is: ", "phonemes of the given speech is: ".
            \item "Descriptive text of the given speech is: ", "The speaking style is: ", "Speech caption is: ".
        \end{itemize}
        \item Ensure the "output" field contains the substring \texttt{|SOA|>audio<|EOA|} and the content that follows it, preserving both the structure and meaning.         \item You may optionally introduce phrases before \texttt{|SOA|>audio<|EOA|} (e.g., "Generated speech is:", "Audio output:", "The resulting audio is:"). Avoid altering or introducing inconsistencies in the audio filename (e.g., \texttt{|SOA|>13\_LibriTTS-R\_260\_123288\_000009\_000000<|EOA|}).
    \end{itemize}

    \item \textbf{Ensure JSON Validity}:
    \begin{itemize}
        \item All strings must be enclosed in double quotes.
        \item Key-value pairs must be separated by commas.
        \item The JSON structure for "instruction", "input", and "output" fields must be valid and consistent.
    \end{itemize}

    \item \textbf{The number of objects in the output should match the number of inputs.} Ensure the format is valid JSON and all JSON objects are properly enclosed. Each entry should be separated by a newline and all JSON should be enclosed in an array.
\end{enumerate}

Please provide the output in valid JSON format (an array of JSON objects), and ensure proper formatting.
\end{tcolorbox}

\subsection{Instruction Variation}\label{appendix:instruction_variation_example}

After the instances (shown in Appendix~\ref{appendix:instruction_template}) are diversified by the prompt in Appendix~\ref{appendix:instruction_variation_prompt}, we obtain instances with various instructions, which are shown below.

\begin{tcolorbox}[colback=white, colframe=black, boxrule=0.2mm, arc=0mm, title=Speech-to-Text Translation Task Template]
\begin{lstlisting}[basicstyle=\ttfamily, breaklines=true]
{
"instruction": "Please translate the speech into the text in English.", 
"input": "<|SOA|>Speech_Audio<|EOA|>", 
"output": "Nevertheless, there are many distinctive ways of drinking coffee around the world that are worth experiencing."
}
\end{lstlisting}
\end{tcolorbox}

\begin{tcolorbox}[colback=white, colframe=black, boxrule=0.2mm, arc=0mm, title=Speech-to-Text Translation Insutrction Variation Example 1]
\begin{lstlisting}[basicstyle=\ttfamily, breaklines=true]
{
"instruction": "Would you be able to convert the spoken words into English text?", 
"input": "<|SOA|>Speech_Audio<|EOA|>", 
"output": "Nevertheless, there are many distinctive ways of drinking coffee around the world that are worth experiencing.", 
}
\end{lstlisting}
\end{tcolorbox}

\begin{tcolorbox}[colback=white, colframe=black, boxrule=0.2mm, arc=0mm, title=Speech-to-Text Translation Insutrction Variation Example 2]
\begin{lstlisting}[basicstyle=\ttfamily, breaklines=true]
{
"instruction": "How about translating the speech into Mandarin English text?", 
"input": "<|SOA|>Speech_Audio<|EOA|>", 
"output": "Nevertheless, there are many distinctive ways of drinking coffee around the world that are worth experiencing.", 
}
\end{lstlisting}
\end{tcolorbox}

\begin{tcolorbox}[colback=white, colframe=black, boxrule=0.2mm, arc=0mm, title=Speech-to-Text Translation Insutrction Variation Example 3]
\begin{lstlisting}[basicstyle=\ttfamily, breaklines=true]
{
"instruction": Please provide the English translation of the audio speech.", 
"input": "<|SOA|>Speech_Audio<|EOA|>", 
"output": "Nevertheless, there are many distinctive ways of drinking coffee around the world that are worth experiencing.", 
}
\end{lstlisting}
\end{tcolorbox}

\begin{tcolorbox}[colback=white, colframe=black, boxrule=0.2mm, arc=0mm, title=Speech-to-Text Translation Insutrction Variation Example 4]
\begin{lstlisting}[basicstyle=\ttfamily, breaklines=true]
{
"instruction": Could you kindly translate the given speech into written English?", 
"input": "<|SOA|>Speech_Audio<|EOA|>", 
"output": "Nevertheless, there are many distinctive ways of drinking coffee around the world that are worth experiencing.", 
}
\end{lstlisting}
\end{tcolorbox}

\begin{tcolorbox}[colback=white, colframe=black, boxrule=0.2mm, arc=0mm, title=Speech-to-Text Translation Insutrction Variation Example 5]
\begin{lstlisting}[basicstyle=\ttfamily, breaklines=true]
{
"instruction": Please provide the English translation of the audio speech.", 
"input": "<|SOA|>Speech_Audio<|EOA|>", 
"output": "Nevertheless, there are many distinctive ways of drinking coffee around the world that are worth experiencing.", 
}
\end{lstlisting}
\end{tcolorbox}

\end{document}